\documentclass[12pt,a4paper]{article}
\usepackage{amsmath}
\usepackage{graphicx}
\usepackage{times}
\usepackage{soul}
\usepackage{cite}
\usepackage{amssymb}
\usepackage{lineno}
\begin{document}
\begin{titlepage}

\title{The invariance of inelastic overlap function}
\author{ S.M. Troshin, N.E. Tyurin\\[1ex]
\small   NRC ``Kurchatov Institute''--IHEP\\
\small   Protvino, 142281, Russia\\}

\normalsize

\date{}
\maketitle

\begin{abstract}
We consider  the symmetry property of the inelastic overlap function and   its relation to the reflective scattering mode appearance. 

\end{abstract}
\end{titlepage}
\setcounter{page}{2}
\section*{Introduction: unitarity of partial   amplitude}
Unitarity written for the partial wave amplitude $f_l(s)$ of elastic scattering (spin degrees of freedom are neglected) has familiar form at high energies, i.e.
\begin{equation}
	\mbox{Im}f_l(s)=|f_l(s)|^2+\eta_l(s),
\end{equation}
where the function $\eta_l(s)$ represents contribution of the intermidiate inelastic states into the product $SS^+$ ($SS^+=1$). Respective elastic  S--matrix element is $S_l(s)=1+2if_l(s)$.
	
Unitarity in the impact parameter representation ($b=2l/\sqrt{s}$) connects the elastic and inelastic  overlap functions introduced by Van Hove \cite{vh} with $h_{tot}(s,b)\equiv \mbox{Im} f(s,b)$  by the following relation
\begin{equation}
h_{tot}(s,b)=h_{el}(s,b)+h_{inel}(s,b)
\end{equation}
The respective cross--sections are determined by  the integrals	of the functions $h_i$ over $b$:
\begin{equation}
\sigma_i(s)=8\pi\int_0^\infty bdb h_i(s,b)
\end{equation}	
where $i=tot,el,inel$. 

This note is devoted to the inelastic overlap function, $h_{inel}(s,b)$, namely, its energy behavior 
at $b=0$.  Based on  unitarity we  consider this quantity as a function of  elastic scattering amplitude. 
 
 Unitarity allows variation of the scattering amplitude in the interval $0\leq |f|\leq 1$ which covers both the absorptive and reflective scattering modes when $|f|\leq 1/2$ and $|f|>1/2$, respectively. Consideration of the inelastic overlap function provides valuable insight on the nature of the scattering.  Transition to the reflective scattering mode results in the peripheral behavior of the inelastic overlap function as well as it changes the structure of the hadron interaction region \cite{07}.

\section{Symmetry  of   inelastic overlap function}
We consider a pure imaginary case of elastic scattering amplitude with replacement $f\to if$. Inelastic overlap function can be expressed then through the scattering amplitude $f(s,b)$ in the following form due to unitarity
\begin{equation}\label{inel}
	h_{inel}(s,b)= f(s,b)[1-f(s,b)].
	\end{equation}
To  clarify the symmetry property of inelastic overlap function consider energy variation of the scattering amplitude $f$ under fixed value of the impact parameter. It is enough to consider the case of $b=0$.
The region of variation of the scattering amplitude covers the range $0\leq f \leq 1$ and $h_{inel}\equiv h_{inel}(s,0) $ invariant under  replacement 
\begin{equation}\label{sym}
f\leftrightarrow1-f.
\end{equation}
Thus, the Eq. (\ref{inel}) is invariant for the two amplitude variation intervals (0, 1/2] and [1/2, 1), i.e.
\begin{equation}\label{symin}
	(0, 1/2]\leftrightarrow [1/2, 1)
\end{equation}
and  the both ranges of amplitude variation correspond to a single range of inelastic overlap function variation $(0,1/4]$.  These two intervals are equivalent in the sense that $h_{inel}$ repeats its values. 
Note, that the use of $U$--matrix unitarization \cite{jpg22} provides a continuous transition from absorptive to reflective scattering mode covering  the whole range of the amplitude variation allowed by unitarity.

The energy evolution of the elastic $S$--matrix scattering element  $S$ ($S\equiv S(s,0)$), the elastic  scattering amplitude $f$ ($f\equiv f(s,0)$) and the inelastic overlap function   from some initial energy $s_i$ to a final value $s_f$
across  the energy $s_m$ where a $h_{inel}$ has its maximal value, (  $h_{inel}^m= 1/4$):
\begin{equation}\label{path}
s_i\to s_{m}\to s_f
\end{equation}
has been discussed in \cite{jpg22}. It is illustrated by the  following relations:
\begin{equation}\label{sii}
	S^i>0 \to S^f<0, \,\mbox{i.e.}\,
f^i<1/2\to f^f>1/2,
\end{equation} and 
\begin{equation}\label{si}
	h_{inel}^i<1/4
	\to h_{inel}^f<1/4.
\end{equation}
 Thus, the inelastic overlap function $h_{inel}$ can perform  a loop variation  with increasing energy  in accordance with Eq.(\ref{path}).  It varies as 
 \begin{equation}\label{hhh}
 	h_{inel}^i\to 1/4\to h_{inel}^f  
 \end{equation}	
and $h_{inel}^f=h_{inel}^i$
 	 provided that 
 $ f^i+f^f=1.$

 Eq. (\ref{si})  means the appearance of the reflective scattering mode and it takes place regardless of the scattering amplitude form.
  Indeed, such behavior of $h_{inel}$ 
 implies an  onset of decrease (i.e. $\partial h_{inel}(s,b)/\partial s$ at $b=0$ becomes  negative) of the inelastic overlap function and is 
associated with appearance of  reflection at the LHC energy range\footnote{The  $U$-matrix form of unitarization and its relation to the symmery property of  $h_{inel}$ was used for continuous extrapolation to higher energies in \cite{jpg22}. It incorporates both scattering modes providing a ground for their simultaneous presence.}.  Impact parameter profile of $h_{inel}(s,b)$ becomes peripheral when $s>s_m$, i.e. $\partial h_{inel}(s,b)/\partial b$ at $b=0$ is strictly positive. Thus,  this invariance of  the inelastic overlap function
 $h_{inel}$
  is in favor of coexistence of the absorptive and reflective scattering modes in elastic scattering (see for discussion of the latter in \cite{07,jpg22}) and corresponds to the transition of the elastic scattering matrix element:
  \begin{equation} \label{refl}
  S\leftrightarrow -S.
  \end{equation} 
 
 It is not clear whether this property has a meaning  of  a separate physical concept.
However,  it is coherent with saturation of  unitarity limit for the amplitude $f$  implied by  the principle of maximal strength of strong interactions proposed long ago by Chew and Frautchi \cite{chew, chew1}. They noted that a `` characteristic of strong interactions is a capacity to `` saturate unitarity condition at high energies''.
   Factor $-1$ in Eq. (\ref{refl}) is interpreted as a result of a reflection by analogy with optics. It can also be considered as a result of an analogue of Berry phase appearance \cite{07} or of color--conducting  matter formation in the interaction region of reaction \cite{jpg}. The color--conducting matter formation can be used in its turn for   explanation of the correspondence of  above symmetry property to a quark--hadron duality (confinement) \cite{brd}. 
   For discussion   of  correlation of the $S$--matrix unitarity and confinement see \cite{np}.  It is also suggested  to associate this mode  with effective resonance formation resulting from  `` an exceptional intermidiate state that unites correlated partons''  \cite{anis,ani,nek}.
  \section{Real part of the scattering amplitude}
   It should be recollected that the above symmetry property for the inelastic overlap function takes place for the pure imaginary scattering amplitude. 
   
   This section discusses the changes in the symmetry properties due to the real part of the scattering amplitude. Relaxing requirement of a pure imaginary  elastic scattering amplitude and taking $b=0$, unitarity condition in the impact parameter representation can be rewritten in the form:
   \begin{equation}\label{inelt}
   	\mbox{Im}f[1-\mbox{Im}f]=h_{inel}+[\mbox{Re}f]^2.
   \end{equation}
It is evident from Eq. (\ref{inelt}) that
\begin{equation}\label{ineq}
[\mbox{Re}f]^2 \leq 1/4-h_{inel}.
\end{equation}
Thus, $|\mbox{Re}f| \to 0$ in both cases: when $h_{inel}\to 1/4$ (full absorption, $S=0$) and/or $\mbox{Im}f\to 1$ (full reflection, $S=-1$). Without neglect of the small real part of the scattering amplitude, one should consider invariance of the function $h_{inel}+[\mbox{Re}f]^2$ under replacement
\begin{equation}\label{symr}
	\mbox{Im}f\leftrightarrow1-\mbox{Im}f.
\end{equation}
Account of a small real part of the scattering amplitude makes picture less transparent but does not change it qualitatively\footnote{Numerical calculations based on the model--independent analysis of available experimental data \cite{tamas} give $h_{inel}=1/4-\alpha$ with  small positive  function $\alpha$, $\alpha\ll 1/4$, at the LHC energies.}. 
 \section*{Conclusion}
 The  symmetry property of the inelastic overlap function and simultaneous presence of the two scattering modes is a consequence of unitarity. It is in favor of coexistence of the absorptive and reflective scattering modes at small impact parameter values\footnote{When addressing the asymptotics with the use of the respective relations, for example, Gribov--Froissart projection formula \cite{grb,frs}, one should not expect the symmetry realization since the scattering is reduced to  purely absorptive mode at large impact parameter values.}. In contrast, the considered symmetry property  disfavors an {\it ad hoc} exclusion of one of  the scattering modes (i.e the reflective  mode) under approaching the asymptotic limit $s\to\infty$. Predominance of the particular  mode is to be correlated with the energy and impact parameters ranges under consideration.

\end{document}